# Interpretable inverse design of optical multilayer thin films based on extended neural adjoint and regression activation mapping

Sungjun Kim and Jungho Kim[*]

*Department of Information Display, Kyung Hee University, 26 Kyungheedae-ro, Dongdaemun-gu, Seoul 02447, Republic of Korea*

[*] Corresponding author.
*E-mail addresses*: sungjunkim@khu.ac.kr (S. Kim), junghokim@khu.ac.kr (J. Kim).

**Abstract:** We propose an extended neural adjoint (ENA) framework, which meets six key criteria for artificial intelligence-assisted inverse design of optical multilayer thin films (OMTs): accuracy, efficiency, diversity, scalability, flexibility, and interpretability. To enhance the scalability of the existing neural adjoint method, we present a novel forward neural network architecture for OMTs and introduce a material loss function into the existing neural adjoint loss function, facilitating the exploration of material configurations of OMTs. Furthermore, we present the detailed formulation of the regression activation mapping for the presented forward neural network architecture (F-RAM), a feature visualization method aimed at improving interpretability. We validated the efficacy of the material loss by conducting an ablation study, where each component of the loss function is systematically removed and evaluated. The results indicated that the inclusion of the material loss significantly improves accuracy and diversity. To substantiate the performance of the ENA-based inverse design, we compared it against the residual network-based global optimization network (Res-GLOnet). The ENA yielded the OMT solutions of an inverse design with higher accuracy and better diversity compared to the Res-GLOnet. To demonstrate the interpretability, we applied F-RAM to diverse OMT structures with similar optical properties, obtained by the proposed ENA method. We showed that distributions of feature importance for various OMT structures exhibiting analogous optical properties are consistent, despite variations in material configurations, layer number, and thicknesses. Furthermore, we demonstrate the flexibility of the ENA method by restricting the initial layer of OMTs to $SiO_2$ and 100 nm.

## 1. Introduction

Optical multilayer thin films (OMTs) are essential components of photonic devices and have a wide range of applications including absorbers (Yang et al., 2016), reflective filters (Fink et al., 1998), antireflection coatings (Guo et al., 2021; Uzum et al., 2017), displays (Bae et al., 2019; Tang and Van Slyke, 1987), solar cells (Britt and Ferekides, 1993; Lee et al., 2021; Repins et al., 2008), and lasers (Faist et al., 1994; Kim and Kim, 2021). Traditionally, the design of OMTs has relied on empirical knowledge. Nevertheless, as the complexity of the design parameters increases, these methods often become trapped in local minima. Consequently, inverse design methodologies have been intensively investigated as a vital field aimed at obtaining the OMT structures with desired optical properties, thereby reducing dependence on human expertise and lowering associated costs. Although previous optimization-based inverse design methods have successfully produced optical structures with the desired optical properties, these approaches are time-consuming because they require numerous estimations of OMTs (Azunre et al., 2019; Rabady and Ababneh, 2014; Schubert et al., 2008; Shi et al., 2017; Tikhonravov et al., 2007).

The rapid computation capabilities of deep neural networks have the potential to significantly reduce the time required for inverse design. Thus, artificial intelligence methodologies have recently emerged as a promising approach for inverse design. Artificial intelligence-assisted inverse design (AIAID) must address the non-uniqueness problem to facilitate the design of OMTs. Neglecting this challenge may compromise the performance of the AIAID algorithm. Key criteria for a sophisticated AIAID methodology include accuracy, efficiency, diversity, scalability, flexibility, and interpretability. The first three criteria ensure that the methodology can provide accurate and diverse optical structures within a reasonable time, thereby enabling engineers to select practical designs (Ma et al., 2024; Tikhonravov et al., 2007; Wang and Guo, 2022). The optical properties of OMTs are influenced not only by the material configuration, layer number, and thickness combination but also by the polarization and incident angle of light. Scalability in the AIAID approach necessitates consideration of all factors that affect optical properties, while flexibility allows for implementation of inverse design within user-defined constraints (Ma et al., 2024). To promote broad exploration, scalability is imperative for an advanced AIAID method. Finally, the interpretability of the model, which enhances the understanding of engineers for deep neural networks through feature visualization, is closely related to the reliability of the model. Consequently, it is necessary to develop an advanced AIAID methodology that meets these six critical requirements.

The straightforward concept of AIAID involves harnessing the forward neural network and the inverse neural network. The forward neural network takes optical structures as input and produces corresponding optical properties as output, while the inverse neural network operates in the reverse manner. The successful learning of the inverse neural network enables inverse design. However, the accuracy of this inverse neural network is degraded due to the non-uniqueness problem for OMTs (Liu et al., 2018). To address this challenge, the tandem neural network, which sequentially connects the inverse neural network with the trained forward neural network, compares actual optical properties with predicted values (Liu et al., 2018). While this tandem approach can generate an accurate optical structure with the desired optical properties, it fails to provide multiple viable design alternatives. In contrast, the neural adjoint (NA) method leverages the trained forward neural network as an accurate interpolator, optimizing optical structures via its backpropagation (Ren et al., 2020). This NA method has produced a variety of structures with the desired optical properties within a reasonable computation time. Nonetheless, the NA method has primarily concentrated on thickness adjustments and has not explored avenues for improving interpretability. These issues can be tackled by improving the forward neural network. Although efforts have been made to enhance the scalability of the forward neural network by perceiving the material and thickness as a word, this has not been discussed within the context of AIAID methodologies (Ma et al., 2023).

The non-uniqueness problem for OMTs can be treated by introducing a probabilistic framework. A notable benefit of probabilistic-based AIAIDs is the ability to generate diverse solutions. Generative models such as the conditional generative adversarial network and conditional variational autoencoder have been employed in the inverse design of OMTs (Dai et al., 2022; Kaireh-Walieh et al., 2023; Kumar et al., 2024). While both models are capable of producing diverse structures, their primary focus has been on adjusting thickness. Neural particle swarm optimization tackles the inverse design of structural color in material configurations and thicknesses by utilizing a mixture density network in conjunction with particle swarm optimization (Wang and Guo, 2022). However, this method remains unexamined concerning layer number for OMTs.

To establish an advanced AIAID methodology, it is essential to achieve scalability. The residual network-based global optimization network (Res-GLOnet) was recently applied to the inverse design for OMTs (Jiang and Fan, 2020). This model enables the inverse design of OMTs with fixed layer number, incorporating arbitrary materials, thicknesses, polarization, and incident angle. Additionally, the layer number of OMTs can also be integrated through iterative computations of Res-GLOnet (Jiang and Fan, 2020). However, this leads to diminished efficiency, and this approach has challenges related to flexibility. In contrast, opto-generative pretrained transformer (OptoGPT) and optical multilayer proximal policy optimization (OML-PPO) secure the scalability to consider the optical properties across arbitrary material configurations, layer numbers, and thicknesses as well as the polarization and incident angle of the light (Ma et al., 2024; Wang et al., 2021). Both models commonly address the inverse design problem for OMTs via sequence modeling and exhibit remarkable performance in terms of accuracy, diversity, scalability, and flexibility. However, the additional numerical optimization is required due to the thickness resolution issue, which furthermore compromises efficiency. However, to the best of our knowledge, a comprehensive AIAID methodology that fulfills six specific requirements has yet to be developed.

In this paper, we propose an extended neural adjoint (ENA) framework, which serves as an AIAID methodology that meets six essential requirements. The ENA represents an advanced version of the previous NA method by introducing the material loss for exploring material configurations. To improve scalability in the previous NA approach, we introduce a novel forward neural network architecture specifically designed for OMTs characterized by arbitrary materials, layer numbers, and thicknesses, referred to as OMT-FNN. The trained OMT-FNN not only provides precise predictions of optical properties across arbitrary material configurations, layer numbers, and thicknesses but also accommodates variations in polarizations and incident angles of light through the application of transfer learning. To enhance the interpretability of the OMT-FNN, we present a specific formulation of regression activation mapping for the OMT-FNN, referred to as F-RAM. The F-RAM, as a feature visualization method, enables the estimation of feature importance. We conduct an ablation study to validate the effect of the material loss introduced in the ENA method, demonstrating that its inclusion improves both diversity and accuracy. The proposed ENA algorithm is also assessed against the Res-GLOnet in terms of accuracy, efficiency, and diversity. Our results show that the ENA method outperforms the Res-GLOnet across all three metrics. Furthermore, we apply F-RAM to diverse OMT structures with similar optical properties, designed by the ENA method. Consequently, our finding indicates that the trained OMT-FNN assigns nearly identical distributions of feature importance to various OMT structures that possess analogous optical properties, despite structural differences present in the target structure and the OMT structures obtained from the ENA method. Finally, we implement the ENA method by constraining the initial layer of OMTs to $SiO_2$ with a thickness of 100 nm, illustrating the flexibility of the proposed ENA method.

- We propose the ENA method, an extended version of NA that further incorporates arbitrary material configurations and layer numbers in the inverse design. We introduce a material loss into the previous NA loss to explore diverse material configurations. A comparative study is conducted between the ENA and the Res-GLOnet algorithm. We demonstrate that the ENA method is an advanced AIAID methodology that fulfills six essential criteria.
- We propose the OMT-FNN architecture, which demonstrates enhanced accuracy, efficiency, scalability, and interpretability. The proposed OMT-FNN is capable of accurately estimating the optical properties for arbitrary materials, layer numbers, and thicknesses. Moreover, it allows for the integration of optical properties considering polarizations and incident angles of light through transfer learning.
- We present a formulation of the F-RAM as a means to improve the interpretability of the OMT-FNN. The F-RAM serves as a feature visualization method that assesses feature importance. This advancement contributes to the enhancement of the overall reliability of the ENA method as well as the OMT-FNN. We demonstrate that the trained OMT-FNN assigns comparable feature importance to OMT structures that exhibit analogous optical properties by utilizing the F-RAM method.

The subsequent sections of this paper are organized as follows: Section 1.1 provides a comprehensive description of the dataset. In Section 2, we review the related work. Section 3 presents the proposed OMT-FNN architecture, the ENA method, the formulation of the F-RAM, and the pseudo-code of the ENA method. Section 4 details the experiment setup and experiment results. Section 5 provides a conclusion with a summary of our research.

*1.1 Dataset description*

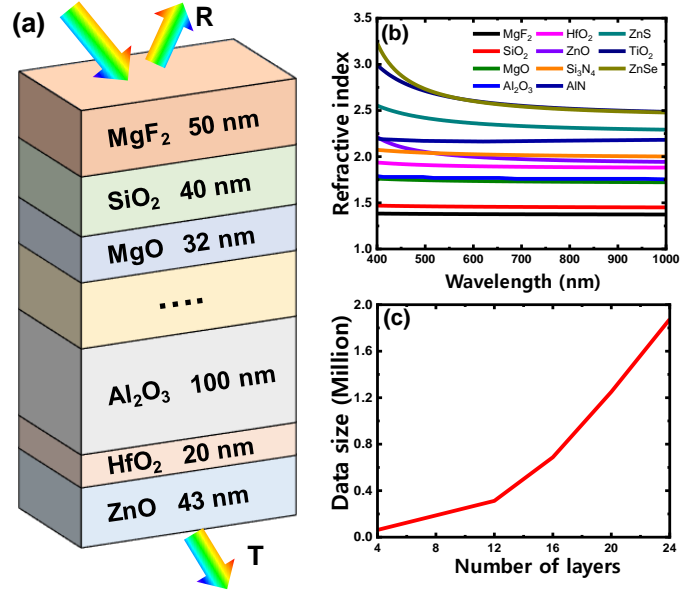

Fig. 1. (a) Schematic diagrams for any OMT structure, where R and T indicate the reflectance and transmittance, respectively. (b) Wavelength-dependent refractive indices of the used dielectric materials, and (c) data size according to layer numbers.

Figure 1(a) illustrates an arbitrary OMT structure. The optical properties of OMTs are influenced by several factors, including the material configuration, layer number, thicknesses, polarization, and the incident angle of incoming light. The optical properties of OMTs are primarily characterized by their reflectance and transmittance. In the context of materials that do not exhibit absorption, these properties can be represented by either transmittance or reflectance. It is essential to take into account the influence of materials characterized by wavelength-dependent refractive indices, as well as their layer thicknesses and numbers, on optical interferences. The optical properties, considering these factors, can be obtained by the transfer matrix method (TMM) simulator.

Figure 1(b) shows the dispersions of dielectric materials used for calculation, of which material indices are summarized in the Supplement. We generated 4,375,000 OMT data for normal incidence from the fourth to the twenty-fourth layers at a four-layer interval without the repetition of materials between adjacent layers. It corresponds to approximately $1/10^{64}$ of the potential structural configurations. Additionally, 437,500 OMT data points are produced for each combination of s and p polarizations at oblique incidences of 20°, 40°, and 60°. A specified range of thicknesses of OMTs is from 20 nm to 100 nm, utilizing an interval of 1 nm. The range of the wavelength is from 400 nm to 1000 nm. Transmittances with a vector size of 301 are extracted as optical properties. Figure 1(c) shows the data size according to layer numbers. We used the TMM simulator provided by (Luce et al., 2022). The dataset generation takes about 37.2 hours on a personal computer with an Intel Core i7-8700 K CPU (3.70 GHz). The total dataset is split into training, validation, and test datasets, following the ratio of 8:1:1.

## 2. Related work

The conventional design of OMTs, which has relied on physical knowledge and intuition, often results in suboptimal outcomes due to the presence of local minima. To mitigate dependence on human experts, significant research has been dedicated to inverse design methodologies for OMTs. Optimization-based inverse design techniques have shown commendable performance in terms of accuracy (Rabady and Ababneh, 2014; Schubert et al., 2008; Shi et al., 2017; Tikhonravov et al., 2007). Furthermore, research has expanded beyond existing thickness optimization (Rabady and Ababneh, 2014) to include investigations into the layer numbers of OMTs (Tikhonravov et al., 2007) and the incorporation of various material configurations (Azunre et al., 2019; Schubert et al., 2008; Shi et al., 2017) to enhance scalability. Nevertheless, these approaches are computationally expensive, necessitating a substantial number of simulations of OMTs.

(Peurifoy et al., 2018) have employed deep neural networks for approximating the nanoparticle dataset, specifically utilizing a forward neural network. This has led to extensive exploration of AIAID methods of photonic devices. To replace previous simulators with the forward neural network, accuracy, efficiency, and scalability are salient. To improve the accuracy of the forward neural network, investigations have been conducted on methodologies such as transfer learning (Qu et al, 2019), active learning (Hong and Nicholls, 2022), and semi-supervised learning (Kim and Kim, 2025). Furthermore, transformer architectures, which leverage attention algorithms, have demonstrated striking performance in natural language processing (Vaswani et al., 2017) and image analysis (Dosovitskiy et al., 2020). Especially, transformers and their hybrid models have been applied to regression tasks (Pokhrel et al, 2022; Song et al, 2025; Yang et al, 2023). The OL-Transformer, a forward neural network based on a transformer encoder, deals with the material and thickness of OMTs as a word (Ma et al., 2023). This model makes precise predictions of optical properties across arbitrary materials, layer numbers, and thicknesses within a pre-defined dataset context, even though there is a thickness resolution issue due to perceiving material and thickness as a word.

For a sophisticated AIAID method, improving the model interpretability is related to reliability. Feature visualization methods have helped humans understand decision-making processes. These methods have been intensively investigated for convolutional neural networks in the image domain (Yosinski et al., 2015; Zeiler and Fergus, 2014). As prominent feature visualization methods, there are a class activation mapping in the classification task (Zhou et al., 2016) and a regression activation mapping in the regression task (Wang and Yang, 2018). The improved interpretability due to these methods led to image localization and discerning feature importance (Gomes and Melo-Pinto, 2021; Wang et al., 2017; Zhou et al., 2018). Although the OL-Transformer visualized the correlation relationship between OMT layers through an attention map (Ma et al., 2023), the suboptimal efficiency of the transformer-based models necessitates the adoption of a streamlined architecture for forward neural networks. However, existing methods for feature visualization have been focused on their applicability from the image domain to the convolutional neural networks. Consequently, there is a need for a methodology to improve the interpretability of forward neural networks that utilize these simplified architectures.

The typical AIAID methodology employing the trained forward neural network is to optimize the design loss to achieve the desired optical properties (Peurifoy et al., 2018). Nonetheless, this approach is characterized by instability owing to a dearth of extrapolation ability of the deep neural network. In contrast, the NA enhanced its stability through incorporating additional boundary loss for thickness, and it has demonstrated remarkable improvements in accuracy, efficiency, and diversity through systematic comparison with other algorithms (Ren et al., 2020). The NA has been recently applied to the inverse design of OMTs, yielding favorable results (Ren et al., 2022; Zhang et al., 2021). Nevertheless, current NA methods exhibit limitations in scalability and flexibility, since they primarily concentrate on the tailoring of thicknesses.

Although the straightforward AIAID method is the direct learning of the inverse neural network, the non-uniqueness problem results in its performance deterioration. Integrating additional optical properties according to the polarizations and incident angles of light into the input of the inverse neural network can mitigate this issue (Lininger et al., 2021). However, the performance degradation becomes increasingly pronounced as the layer number increases. Tandem neural network treats the non-uniqueness problem by comparing actual optical properties with predicted values (Liu et al., 2018) and has focused thicknesses of OMTs (Kim et al., 2021; Xu et al., 2021). Recent studies have sought to enhance accuracy by changing the model architecture (Chen et al., 2023) and to improve scalability by incorporating additional loss functions (So et al., 2019). Nevertheless, the tandem approach provides only one optical structure, and training two deep neural networks may compromise overall efficiency.

The non-uniqueness problem can be tackled through probabilistic frameworks. Conditional generative adversarial networks and conditional variational autoencoders have been applied to the inverse design of OMTs (Dai et al., 2022; Kaireh-Walieh et al., 2023; Kumar et al., 2024). Although conditional generative adversarial networks-based inverse design provides various optical structures, they may encounter stability issues, such as mode collapse. Conditional variational autoencoder-based inverse design exhibits good robustness, although it tends to lack diversity due to a narrow distribution of optical structures (Jang and Kim, 2024). A mixture density network defines the output of the inverse neural network as mean values, variances, and weights of multiple Gaussian distributions, procuring the diversity (Unni et al., 2020). In the context of material configurations, the neural particle swarm opimization utilizes the mixture density network to select material configurations and thicknesses, while recommended thicknesses of OMTs are subsequently finetuned through the particle swarm optimization (Wang and Guo, 2022). Although this model shows good performance for the inverse design of OMTs, it is limited to scenarios involving a fixed layer number.

The global optimization network optimizes the optical properties of optical devices by exploiting both the forward and adjoint computations of the electromagnetic simulators, such as rigorous coupled wave analysis (Jiang and Fan, 2019). This trained network produces diverse optical structures with optimized performance. Notably, an advantage of this model is that it does not necessitate a pre-established dataset. Nevertheless, the forward and adjoint computations of simulators can be computationally intensive. The Res-GLOnet treats the thicknesses and material configurations for OMTs with a fixed layer number, showing a great performance (Jiang and Fan, 2020). The Res-GLOnet employs a TMM simulator developed using the PyTorch framework, which substitutes the adjoint computation with automatic differentiation. This approach markedly improves efficiency compared to the previous global optimization network, as the TMM simulator operates at a much faster rate than traditional electromagnetic simulators. While the model accommodates layer numbers by repeating it according to specified layer dimensions, this adaptation results in reduced efficiency. According to (Jiang and Fan, 2020), the Res-GLOnet demonstrated superior accuracy and efficiency compared to the OML-PPO. However, the Res-GLOnet is constrained by a lack of flexibility.

The OptoGPT and the OML-PPO address the inverse design problem for arbitrary materials, layer numbers, and thicknesses through sequence modeling (Ma et al., 2024; Wang et al., 2021). The OptoGPT facilitates the inverse design of precise OMT structures by exploiting self-attention and cross-attention. The self-attention is utilized to learn the relationship within components of the OMT structure, while the cross-attention is implemented between these components and their corresponding optical properties. The probability sampling of this model inherently promotes diversity, and the probability resampling enhances flexibility. The OML-PPO, on the other hand, determines the material and thickness of OMTs in an autoregressive manner. Furthermore, to avoid the repetition of materials across adjacent layers, a non-repetitive gating function was introduced, leading to the prevention of redundant calculations. Notably, the OML-PPO does not rely on a pre-defined dataset and enables the optimization of optical properties due to the characteristics of reinforcement learning. Even though both the OptoGPT and the OML-PPO reported striking performance in inverse design, both models have a thickness resolution, as they perceive thickness as a word. This requires additional numerical optimization for fine-tuning thicknesses. Therefore, there is a need for the development of an advanced AIAID methodology that balances accuracy, efficiency, diversity, scalability, flexibility, and interpretability.

## 3. Methodology

In this chapter, we present the OMT-FNN architecture, which is designed to precisely predict the optical properties of OMTs with various material configurations, different layer numbers, and thicknesses. We outline the ENA framework, allowing for simultaneous optimization with respect to materials, layer numbers, and thicknesses. We present the formulation of the F-RAM aimed at enhancing the interpretability of the model.

*3.1 The forward neural network architecture for optical multilayer thin films*

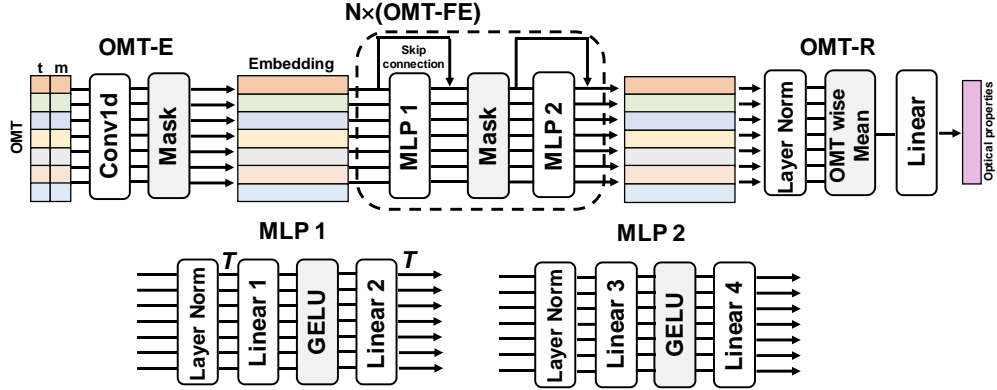

Fig. 2. The proposed OMT-FNN architecture

Figure 2 illustrates the proposed OMT-FNN architecture, which comprises three primary components: the OMT embedding (OMT-E) layer, the OMT feature extractor (OMT-FE) layer, and the OMT regressor (OMT-R) layer. The input of the OMT-E layer consists of the thicknesses and material indices of the OMTs with an arbitrary number of layers. For the sake of clarity, the batch dimension is omitted in this explanation. The thickness and material vectors are denoted as $t \in \mathbb{R}^{d_i \times 1}$ and $m \in \mathbb{Z}^{d_i \times 1}$, respectively, with $d_i$ representing the maximum layer number of the OMTs within the dataset. For OMTs with layer numbers shorter than $d_i$, the regions beyond their respective layer numbers are filled with padding values, utilizing -1 as the padding value. To balance the input of OMTs, min-max normalization is applied to $t$ and $m$. The convolution layer within the OMT-E integrates the material and thickness information of the OMT, while the masking layer effectively blocks the influence of padding values. The output generated by the OMT-E for the OMT is an embedded matrix, denoted as $X^{(e)} \in \mathbb{R}^{d_i \times d_e}$, where $d_e$ is the embedding dimension. The rows of $X^{(e)}$ encapsulate information pertaining to each layer of the OMT, whereas the columns of $X^{(e)}$ convey information regarding the embeddings of the OMT.

The OMT-FE layer, which consists of two multi-layer perceptron (MLP) components, namely MLP 1 and MLP 2, is responsible for transforming the embedding matrix of the OMT into a feature matrix (Tikhonravov, A. V. et al, 2021). The transformation formula is given by

$$O^{(e)} = \left(\sigma\left(LN\left(X^{(e)}\right)^T W_1\right) W_2\right)^T + X^{(e)}, \tag{1}$$

$$F^{(e)} = \sigma\left(LN\left(O^{(e)}\right) W_3\right) W_4 + O^{(e)}, \tag{2}$$

where $T$ denotes transpose, and $\sigma$ indicates the activation function utilized by the Gaussian error linear unit (GELU). The abbreviation *LN* refers to layer normalization. The weight matrices associated with the linear layers of the MLP 1 are denoted as $W_1$ ($\in \mathbb{R}^{d_i \times 2d_i}$) and $W_2$ ($\in \mathbb{R}^{2d_i \times d_i}$), while the corresponding matrices for the MLP 2 are represented as $W_3$ ($\in \mathbb{R}^{d_e \times 2d_e}$) and $W_4$

($\in \mathbb{R}^{2d_e \times d_e}$). The role of MLP 1 is to mix the information of layers of the OMT through the weighted summation, whereas MLP 2 is responsible for mixing the embeddings of the OMT. As a result of the MLP 1, the values are filled in the padding rows. To prevent the influence of these contributions, a masking layer is employed. Although this approach leads to a slight reduction in accuracy, the incorporation of the masking layer improves the interpretability of the model. Therefore, the feature matrix extracted from the OMT-FE represents the integrated information of the OMT layers and their corresponding embeddings. Additionally, a skip connection is introduced to improve the stability in the OMT-FE (He et al., 2016).

The OMT-R layer converts the feature matrix into the prediction of the optical properties of the OMT, as expressed as by the following equation:

$$y = \frac{1}{d_i} \sum_{p=1}^{d_i} LN\left(F_{pq}^{(e)}\right) W_R, \tag{3}$$

where y ($\in \mathbb{R}^{1 \times d_o}$) indicates the predicted optical properties of the OMT. $d_o$ denotes the output dimension. The term $W_R$ ($\in \mathbb{R}^{d_e \times d_o}$) refers to the weight associated with the linear layer of the OMT-R layer, respectively. As shown in Fig. 2, the OMT-wise mean indicates the column-wise mean of the feature matrix.

*3.2 Extended neural adjoint method*

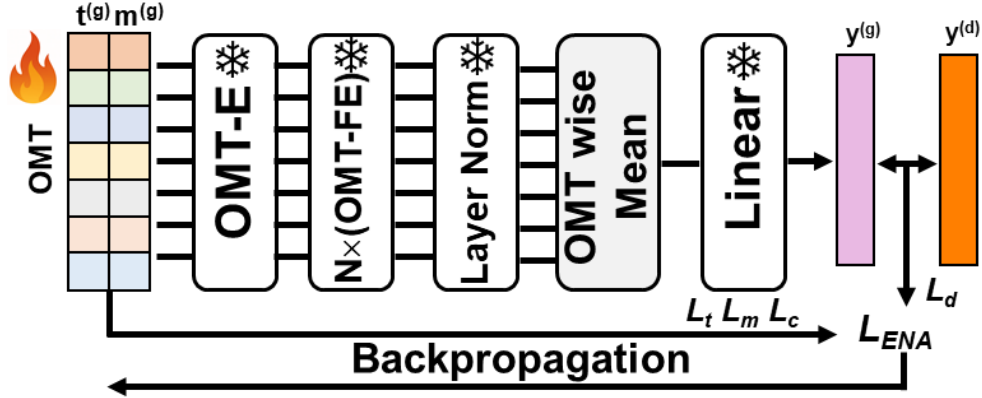

Fig. 3. The proposed ENA process. The flame and snowflake symbols denote learnable and fixed parameters, respectively.

We present the ENA formulation, which facilitates the concurrent optimization of material configurations, layer numbers, and thicknesses. As shown in Fig. 3, the frozen OMT-FNN is used as the estimator for the optical properties. It enables accurate predictions of optical properties of OMTs characterized by various material configurations, different numbers of layers, and diverse thickness combinations. The losses for the previous NA (Ren et al., 2020) are expressed as

$$L_d = \mathbb{E}_g\left[SE\left(y^{(g)}, y^{(d)}\right)\right], \tag{4}$$

$$L_t = \mathbb{E}_g\left[ReLU\left(\left|t^{(g)} - \mu^{(t)}\right| - \frac{1}{2}M^{(t)}\right)\right], \tag{5}$$

where $L_d$ and $L_t$ represent the design loss for the desired optical properties and the boundary loss for the thickness, respectively. $y^{(d)}$ denotes the desired optical properties, and the superscript $g$ indicates generated candidates for OMT. $t^{(g)}$ indicates the generated thickness vectors. $\mu^{(t)}$ and $M^{(t)}$ represent the mean and magnitude vectors of the thickness range. The terms *ReLU* and *SE* stand for rectified linear unit and squared error, respectively. $L_t$ is instrumental in restricting OMT candidates within the specified thickness range. Note that the optimization of (4) and (5), employing the OMT-FNN, implements the inverse design of OMTs by solely modulating the thickness of layers within a predetermined set of diverse material configurations and a various number of layers. This approach is referred to as ENA without material loss, in which $t^{(g)}$ is the only learnable parameter. It corresponds to the improved version of the previous NA, which is limited to adjusting thicknesses within a fixed material configuration and a constant number of layers. However, the limitation in the ability to explore various material configurations of the ENA without material loss constrains its efficacy as an inverse design methodology. Therefore, incorporating material configurations as learnable parameters is essential for enhancing the performance of the NA-based inverse design method.

In order to effectively perform the NA-based inverse design method by integrating thicknesses and material configurations as learnable parameters, it is crucial to adhere to two key requirements. Firstly, the optimization process must ensure that material configurations remain within established bounds. Secondly, it is imperative to avoid material redundancy between adjacent layers. The reason is why presence of configurations that fall outside the specified range or exhibit redundancy may compromise the accuracy of the OMT-FNN. This is due to the fact that the dataset utilized for training the OMT-FNN is distributed within a specific range and is devoid of material repetition between adjacent layers. To account for these requirements, we introduce the following material loss function, which is given by

$$L_m = L_m^{(b)} + L_m^{(r)}, \tag{6}$$

$$L_m^{(b)} = \mathbb{E}_g \left[ ReLU \left( \left| m^{(g)} - \mu^{(m)} \right| - \frac{1}{2} M^{(m)} \right) \right], \tag{7}$$

$$L_m^{(r)} = \mathbb{E}_g \left[ ReLU \left( -\sum_{q=1}^{d_i} \left| m_{q+1}^{(g)} - \lfloor m_q^{(g)} \rfloor \right| + 1 \right) \right]. \tag{8}$$

Here, $m^{(g)}$, which is a learnable parameter, denotes the generated material vectors. $\mu^{(m)}$ and $M^{(m)}$ indicate the mean and magnitude vectors of the material range. The symbol $\lfloor \ \rfloor$ represents the rounding operation. $L_m^{(b)}$ corresponding to boundary loss for the material index serves to suppress the material indices from being out of range. $L_m^{(r)}$, referred to as a material redundancy regularization loss, facilitates the elimination of material repetition. Therefore, the material loss enables exploring diverse material configurations. The subsequent loss function for the ENA is expressed as

$$L_{ENA} = L_d + w_1 L_t + w_2 L_m + w_3 L_c, \tag{9}$$

$$L_c = \mathbb{E}_g \left[ SE \left( t^{(g)}, t^{(d)} \right) \right] + \mathbb{E}_g \left[ SE \left( m^{(g)}, m^{(d)} \right) \right], \tag{10}$$

where $w_p$ is the weighting factor. $L_c$ indicates a constraint loss to enhance the flexibility of the ENA method. They play a role in constraining the layer thickness and material index of the specific layers. We refer to optimizing (9) as an ENA method.

*3.3 Regression activation mapping for the OMT-FNN*

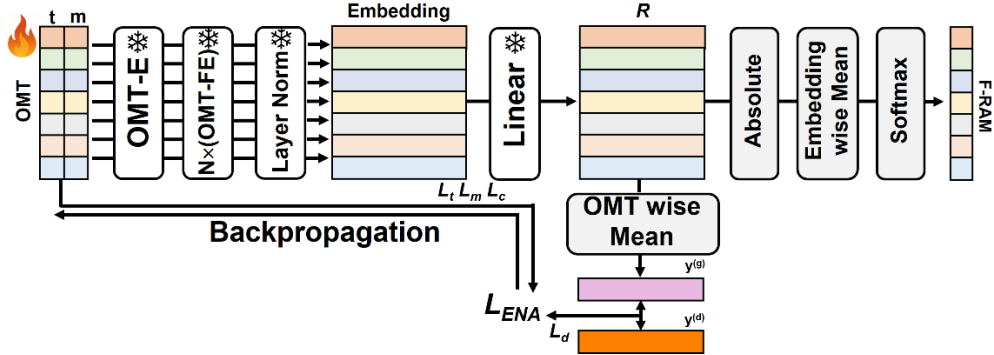

Fig. 4. The proposed F-RAM process. The flame and snowflake symbols denote learnable and fixed parameters, respectively. It is important to not that the computation of feature importance through F-RAM can be exploited in the ENA-based inverse design as well as the forward prediction of OMTs.

To improve the interpretability of the trained OMT-FNN, we implement the F-RAM formulation, which is a feature visualization method. Figure 4 shows the F-RAM process. An OMT feature matrix is generated through sequential application of the OMT-E, OMT-FE, and layer normalization layers for any given OMT. The F-RAM matrix is subsequently derived by utilizing the linear layer of the OMT-R, as detailed below

$$R = LN\left(F^{(e)}\right)W_R, \qquad (11)$$

where $R$ ($\in \mathbb{R}^{d_i \times d_o}$) indicates the F-RAM matrix. The elements of $R$ are represented by the inner product between the feature vectors extracted from any given OMT and the weight vectors of the trained OMT-R. It is important to note that the prediction of the optical properties is achieved through the application of the OMT-wise mean to $R$. This means that $R$ includes information about the decision of the trained OMT-FNN. A positive contribution to the prediction of optical properties arises when the row vector of $F^{(e)}$ is aligned with the column vector of $W_R$. Conversely, the opposite direction between two vectors results in a negative contribution. Moreover, if these vectors are orthogonal, their influence on the prediction is negligible. Thus, the absolute value of the elements of $R$ indicates the feature importance.

The F-RAM is derived by applying the embedding-wise mean, which is row-wise mean, to $|R|$, and it is expressed as

$$r = Softmax\left(\frac{1}{d_e}\sum_{q=1}^{d_e}\left|R_{pq}\right|\right), \qquad (12)$$

where $r$ ($\in \mathbb{R}^{d_i \times 1}$) is the F-RAM. A soft-max function is applied for normalization. Thus, $r$ is analyzed as the importance of the feature vector extracted from the OMT-FE layer, since the components of $r$ express the contribution to the prediction. Note that (12) is similar to the attention score formula (Vaswani et al., 2017). A notable advantage of the F-RAM method is its applicability across various dimension-preserving deep neural network architectures.

*3.4 Pseudo-code of extended neural adjoint algorithm*

The pseudo-code for the ENA algorithm is presented in Algorithm 1. Initially, the parameters of the trained OMT-FNN are fixed, while the thicknesses and material indices of the OMT candidates with the arbitrary layer number are randomly generated. The layer numbers of the initial individuals are created such that they are uniformly sampled from the predefined set of possible layer numbers, which ranges from the minimum number to the maximum number. By generating initial OMT candidates with varying layer numbers and performing the ENA method, we include the influence of layer number as well as material and thickness. Subsequently, the following steps are executed iteratively until the maximum iteration limit is reached. The optical properties of the generated OMT candidates are predicted using the trained OMT-FNN. The design loss is computed based on these predictions and desired optical properties. Additionally, the boundary loss for thickness, material loss, and constraint loss are calculated. The ENA loss is obtained through the weighted summation of these losses. The thicknesses and material indices of the OMT candidates are then updated by optimizing the ENA loss. Following these iterations, the rounding operation is applied to the optimized material indices to ensure integer values. Finally, the optimized OMT structures experience the filtering process to remove candidates that fall outside the range of thickness and material index, as well as those that exhibit material redundancy.

---

**Algorithm 1: Extended neural adjoint algorithm**

**Input**: desired optical properties $y^{(d)}$, trained OMT-FNN parameters $\phi$, the number of OMT candidates $G$, mean vectors and magnitude vectors of thickness and material ranges $\mu^{(t)}$, $\mu^{(m)}$, $M^{(t)}$, and $M^{(m)}$, weights $w_1$, $w_2$, and $w_3$, learning rate $\eta$

**Result**: Optimized OMT solutions

Freeze OMT-FNN parameters
Generate initial OMT candidates: $\{t^{(g)}, m^{(g)}: g = 1, 2, …, G\}$
**for** $t = 0$ to E-1 do
  Compute the forward predictions: $y^{(g)} = f_\phi(t^{(g)}, m^{(g)})$
  Compute the design loss and boundary loss for thicknesses: $L_d$ and $L_t$
  Compute the material loss: $L_m$
  Compute the restriction loss: $L_c$
  Compute the ENA loss: $L_{ENA} = L_d + w_1 L_t + w_2 L_m + w_3 L_r$
  Update OMT candidates: $[t^{(g)}, m^{(g)}] = [t^{(g)}, m^{(g)}] - \eta \nabla_{[t^{(g)}, m^{(g)}]} L_{ENA}$

**End**

Apply the rounding operation to optimized material indices: $\lfloor m^{(g)} \rfloor$

Filter out optimized OMT candidates: Thickness and material index out of range, and material index redundancy

---

## 4. Experiment

### 4.1 Experiment setting

The hyperparameters used in experiments are summarized in Section 1 of the Supplement. The experiments were conducted using an NVIDIA GeForce RTX 4090 GPU.

### 4.1.1 Performance metrics

We harness root mean squared error (RMSE) and determination coefficient ($R^2$ score) as accuracy metrics, while the efficiency metrics are utilized by training time and inverse design time. The formulas of the RMSE and $R^2$ score are represented as

$$RMSE = \sqrt{\frac{1}{d_o} \sum_{j=1}^{d_o} (\hat{y}_j - y_j)^2}, \tag{11}$$

$$R^2 = 1 - \frac{\sum_{j=1}^{d_o} (y_j - \hat{y}_j)^2}{\sum_{j=1}^{d_o} (y_j - \bar{y})^2}, \tag{12}$$

where $y$, $\bar{y}$, and $\hat{y}$ denote the desired optical properties, the mean desired optical properties, and model prediction, respectively. Diversity is quantified by statistical properties of selected individuals $C$, defined as $C = \{c \mid R^2(s) \geq 0.9, s \in S\}$, where $S$ represents the optimized OMT solution set. As statistical metrics, the number of solutions in the chosen set and the average standard deviation of layer-wise thicknesses and material indices across the selected OMT solutions are employed. The formulas for standard deviations are expressed as

$$\sigma_t = \frac{1}{N} \sum_{n=1}^{N} \sqrt{\frac{1}{|C|} \sum_{c \in C} \left( t_n^{(c)} - \bar{t}_n \right)^2}, \tag{13}$$

$$\sigma_m = \frac{1}{N} \sum_{n=1}^{N} \sqrt{\frac{1}{|C|} \sum_{c \in C} \left( m_n^{(c)} - \bar{m}_n \right)^2}, \tag{14}$$

where $|C|$ indicates the cardinality of $C$ and $N$ denotes the layer number. A greater standard deviation indicates a higher level of diversity.

*4.2 Experiment results*

*4.2.1 Performance of the OMT-FNN*

The presented OMT-FNN architecture is trained in a supervised learning manner. The resulting performance metrics for the trained OMT-FNN include an RMSE of 0.010, an $R^2$ score of 0.999, and a total training time of 33.2 hours. To explore effective feature extractor architectures, we conducted a comparative analysis by modifying the feature extractor to the MLP, the convolutional neural network, and the transformer encoder layers. We demonstrate that the proposed OMT-FE architecture outperforms other architectures across accuracy and efficiency. Additionally, we applied transfer learning to improve the scalability of the trained OMT-FNN, using datasets consisting of combinations of polarizations and incidence angles, which is one-tenth the size of the original dataset. As a result, we demonstrate that the performance of the fine-tuned models is commendable. The details are presented in Section 2 of the Supplement. A comparative evaluation of the trained OMT-FNN against the OL-Transformer model (Ma et al., 2023) is also included, with further details provided in Section 3 of the Supplement.

*4.2.2 Performance demonstration of material loss in the ENA method*

To validate the effect of the material loss introduced in the ENA for the purpose of exploring material configurations, we conduct a comparative analysis between the ENA and both the ENA without material loss and the ENA without material redundancy regularization loss, which corresponds to weighted summation of (4), (5), and (7). The ENA without material loss is readily reproduced by setting $w_2$ to zero. We designate one of the test datasets as the target spectrum. The total population is designated as 1,200, which corresponds to 200 populations for each layer number. The methodologies are evaluated based on their accuracy, efficiency, and diversity. In this experiment, we set $w_3$ as 0.

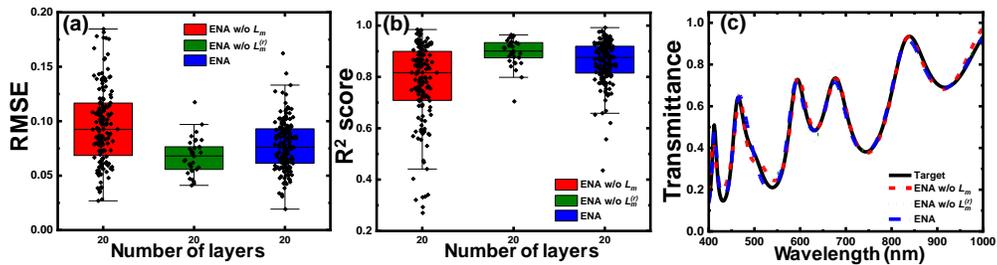

Fig. 5. Box overlap plot for (a) RMSE and (b) $R^2$ score of the OMT solutions with a layer number of 20 obtained using the three AIAID methods, and (c) Transmittance spectra for the target structure and those obtained using the three AIAID methods: ENA without material loss, ENA without material redundancy regularization loss, and ENA.

**Table 1. The performance metrics of the OMT structures derived from ENA without material loss, ENA without material redundancy regularization loss, and ENA methods**

| Model | RMSE | $R^2$ score | $|C|$ | $\sigma_t$ | $\sigma_m$ | Inverse design time (sec) |
|---|---|---|---|---|---|---|
| ENA w/o $L_m$ | 0.027 | 0.985 | 83 | 25.690 | 3.091 | 41.7 |
| ENA w/o $L_m^{(r)}$ | 0.041 | 0.963 | 81 | 23.791 | 2.969 | 48.9 |
| ENA | 0.020 | 0.992 | 165 | 26.423 | 3.202 | 55.9 |

Figure 5(a) and (b) present box overlap plots illustrating RMSE and $R^2$ score for the OMT solutions with a layer number of 20 obtained from the ENA without material loss, the ENA without material redundancy regularization loss, and the ENA. Table S7 of the Supplement delineates the target OMT structure alongside the OMT structures exhibiting the highest accuracy achieved through these three methods, while Table 1 provides a comparative analysis of the performance metrics associated with these methods. Figure 5(c) shows transmittance spectra for the OMT structures listed in Table S7.

The ENA without material loss yields the OMT structure with comparable accuracy to the ENA method, underscoring the significant impact of tailored thickness on the inverse design of OMT. This indicates that, given appropriate material configurations, the performance of the ENA without material loss is great. However, the total number of possible material configurations considered in this study is approximately $10^{24}$, making it difficult to produce suitable material configurations through random generation. The diversity of the OMT solutions generated by the ENA without material redundancy regularization loss exhibits a deterioration in the metrics $|C|$, $\sigma_t$, and $\sigma_m$ in comparison to the ENA without material loss. Additionally, the peak accuracy experiences a decline in the ENA without material redundancy regularization loss. As illustrated in Fig. 5(a) and (b), the total number of OMT solutions produced by the ENA without material redundancy regularization loss is lower than that of the two other methods, with 586 OMT solutions being removed during the filtering process due to material redundancy. In contrast, the ENA method results in a reduction of filtered OMT solutions to 78 due to material redundancy. This observation highlights the efficacy of the inclusion of the material redundancy regularization loss in facilitating the exploration of material configurations. Moreover, the diversity of the ENA method is largely enhanced across $|C|$ compared to the ENA without material loss and the ENA without material redundancy regularization loss. Consequently, not only is the best accuracy of the ENA enhanced, but the diversity of the OMT solutions is also significantly improved, attributable to the comprehensive exploration of material configurations, layer numbers, and thicknesses within the ENA method.

*4.2.3 Performance comparison of the ENA method to the Res-GLOnet*

We focus on a specific inverse design task of a band-pass filter that has a unity transmittance between 600 and 700 nm and a zero value outside this spectral range. We conduct a comparative analysis against the Res-GLOnet (Jiang and Fan, 2020). The Res-GLOnet, consisting of residual network-based generative neural networks, produces OMT structures with a constant layer number characterized by material configurations and thicknesses. The parameters of the Res-GLOnet are updated through the automatic differentiation of the TMM simulator. In this study, we reproduced it using the Pytorch framework. Therefore, the Res-GLOnet has inherent accuracy through the TMM simulator. The hyperparameters of the Res-GLOnet are presented in (Jiang and Fan, 2020). The total epochs and populations in the Res-GLOnet were modified to 2000 and 600, respectively. To ensure a fair comparison, the population size for the ENA method was designated as 3600, which equates to 600 populations per layer number. The Res-GLOnet was executed sequentially at intervals of four layers, from the fourth to the twenty-fourth layers, while the ENA method was implemented in parallel at the same four-layer intervals. In the application of the ENA method, the weighing factors designated as $w_1$, $w_2$, and $w_3$ are assigned values of 1, 1, and 0, respectively.

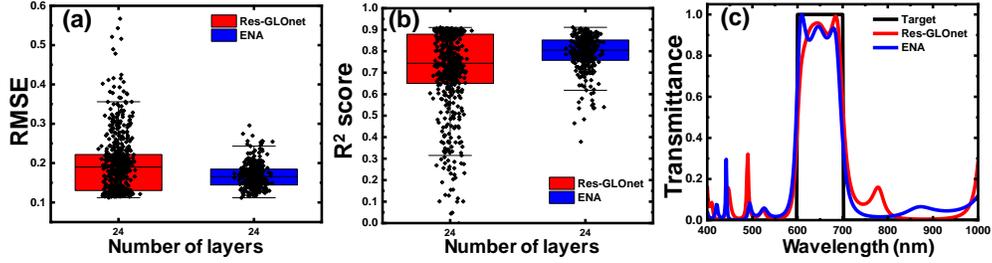

Fig. 6. Box overlap plot for (a) RMSE and (b) $R^2$ score of the OMT solutions with a layer number 24 obtained using Res-GLOnet and ENA method. (c) Transmittance spectra for the target and those obtained using Res-GLOnet and ENA method.

**Table 2. The performance metrics of the OMT structures with a layer number of 24 obtained by the Res-GLOnet and ENA methods**

| Model | RMSE | $R^2$ score | $|C|$ | $\sigma_t$ | $\sigma_m$ | Inverse design time (sec) |
|---|---|---|---|---|---|---|
| Res-GLOnet | 0.114 | 0.907 | 27 | 2.707 | 0.048 | 6,789.5 |
| ENA | 0.111 | 0.912 | 9 | 20.072 | 2.411 | 131.2 |

Figures 6(a) and (b) show box overlap plots for RMSE and $R^2$ score of the OMT solutions with a layer number of 24 obtained by the Res-GLOnet and the ENA method. The OMT structures with the highest accuracy are presented in Table S6 of the Supplement, and corresponding spectra are shown in Figure 6(c). Their performance metrics are presented in Table 2.

In terms of accuracy, the ENA achieves higher performance. The Res-GLOnet and the ENA provide diverse solutions. Specifically, the Res-GLOnet produces 27 OMT structures with a layer number of 24, while the ENA generates 2 and 7 OMT structures with layer numbers of 20 and 24, respectively. The distribution of the OMT structures obtained from the Res-GLOnet and the ENA method is visualized in Section 4 of the Supplement. Although the Res-GLOnet produces greater OMT solutions than the ENA method, the average standard deviations for thickness and material index in the ENA method are 7.4 times and 50.2 times greater than those observed in the Res-GLOnet, respectively. Consequently, the OMT solutions obtained by the ENA method demonstrate a greater diversity in the comparison to the Res-GLOnet. This means that OMT solutions derived from the ENA method are distributed across multiple local minima, while those from the Res-GLOnet are predominantly clustered around a singular local minimum. Therefore, the probability of finding the global minimum in the ENA method is greater than that in the Res-GLOnet.

### 4.2.4 Interpretability demonstration via F-RAM

We illustrate the interpretability of the trained OMT-FNN utilizing F-RAM. F-RAM visualizations for 100 arbitrary selected OMT test data with layer numbers of 4, 8, 12, 16, 20, and 24 are presented in Section 6 of the Supplement. In this section, we elucidate how the trained OMT-FNN addresses the non-uniqueness problem associated with OMTs by applying the F-RAM to the target structure and the selected OMT structures, obtained from the ENA method in Section 4.2.2.

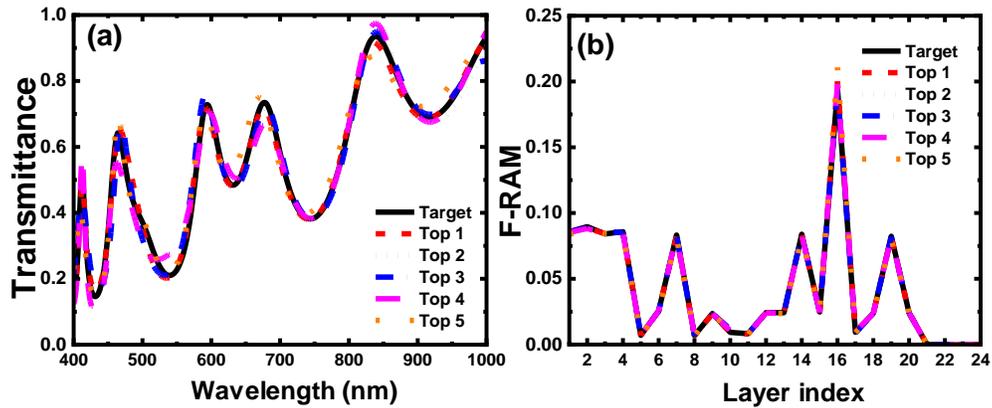

Fig. 7. (a) Transmittance spectra and (b) F-RAMs for the target structure and the five most accurate OMT structures with a layer number of 20, designed using the ENA method.

Table 3. The accuracy metrics of the five most accurate OMT structures with a layer length of 20, designed using the ENA method.

| Rank | RMSE | $R^2$ score |
|---|---|---|
| Top 1 | 0.020 | 0.992 |
| Top 2 | 0.030 | 0.980 |
| Top 3 | 0.033 | 0.976 |
| Top 4 | 0.035 | 0.974 |
| Top 5 | 0.040 | 0.966 |

Table S8 in the Supplement presents the five most accurate OMT structures of a layer number of 20, while the corresponding transmittance spectra and accuracy metrics are shown in Figure 7(a) and Table 3, respectively. Figure 7(b) illustrates F-RAMs for the target structure and the OMT structures listed in Table S8.

As shown in Fig. 7(b), the F-RAMs are nearly identical, despite the differences in thickness and material across the layers of the target structure and the five most accurate OMT structures. This result shows that the trained OMT-FNN assigns nearly equivalent feature importance to OMT structures exhibiting comparable optical properties. This demonstrates that, despite the physical distinctions present among OMT structures with similar optical properties, the trained OMT-FNN clusters these structures based on their feature importance.

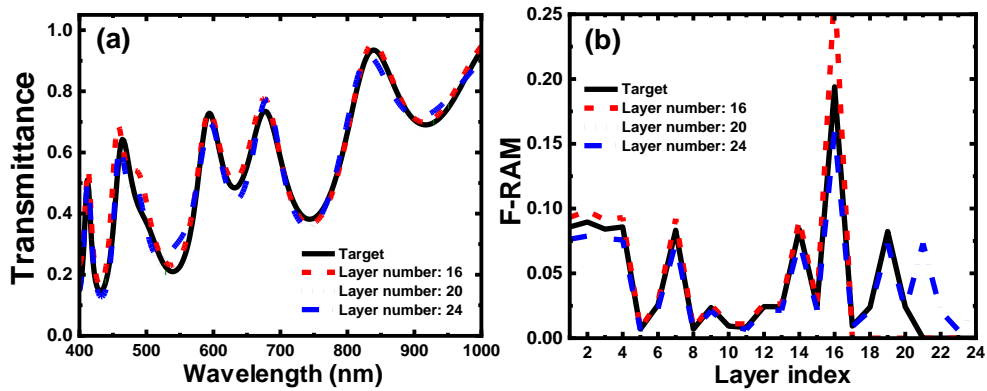

Fig. 8. (a) Transmittance spectra and (b) F-RAMs for the target structure and the OMT structures with layer numbers of 16, 20, and 24, designed using the ENA method.

**Table 4. The performance metrics of the OMT structures with layer lengths of 16, 20, and 24, designed using the ENA method.**

| Layer number | RMSE | $R^2$ score |
|---|---|---|
| 16 | 0.039 | 0.968 |
| 20 | 0.020 | 0.992 |
| 24 | 0.034 | 0.976 |

Table S9 in the Supplement shows the OMT structures with layer numbers of 16, 20, and 24 obtained by the ENA method. These structures are associated with the highest accuracy corresponding to their respective layer numbers, and their accuracy metrics are presented in Table 4. Figure 8(a) and (b) illustrate the transmittance spectra and F-RAMs for both the target structure and the OMT structures listed in Table S9, respectively.

As shown in Fig. 8(b), it is noticeable that F-RAM does not account for contributions from padding rows. Although the specific F-RAMs for the OMT structures with layer numbers of 16 and 24 differ from those of the target structure, they exhibit a similar trend in the graph regarding increases, decreases, or peak positions. These observations from Fig. 7(b) and Fig. 8(b) suggest that, despite differences in material configurations, layer numbers, and thicknesses, the trained OMT-FNN assigns comparable feature importance to OMT structures exhibiting similar spectral characteristics, thereby facilitating pattern recognition within the OMT dataset.

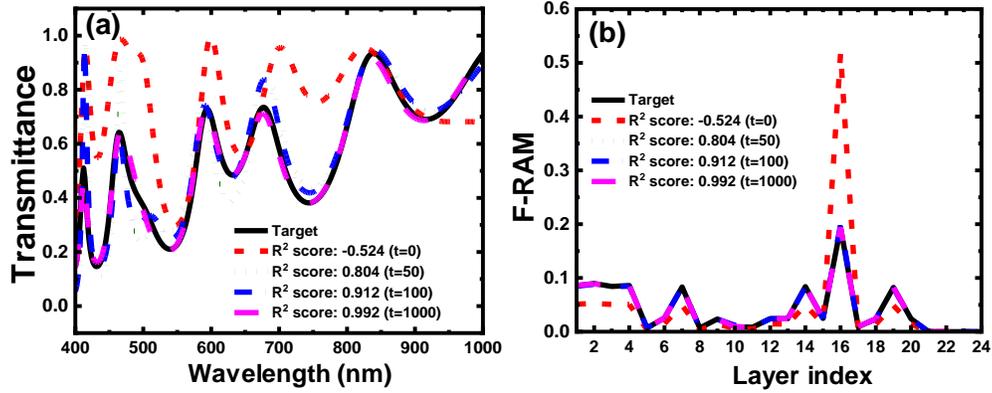

Fig. 9. (a) Transmittance spectra and (b) F-RAMs for the target structure and the OMT structures obtained during the ENA method in the 20-layer OMT solution with the highest accuracy. t denotes the time step of the ENA method

To improve the interpretability of the ENA method, we analyze the progression of the transmittance spectrum and the F-RAM in relation to the revised OMT structure, specifically layer number 20 as outlined in Table S9, during the ENA method. At the time step of 0, the transmittance spectrum and F-RAM display significant discrepancies when compared to those of the target structure, as the OMT structure at this initial stage corresponds to randomly generated thicknesses and materials. After 50 iterations, the $R^2$ score for the updated OMT structure shows an enhancement of 1.328 relative to the time step of 0. As the iteration number increases, the transmittance spectrum approaches the target spectrum, although certain peaks of the transmittance spectrum at the time steps of 50 and 100 still reveal differences. As shown in Fig. 9(b), the feature importance for layer index 16 at the time step of 0 is distributed across subsequent time steps, with the F-RAM converging towards that of the target structure. As the time steps progress towards 1000, both the transmittance spectrum and F-RAM undergo refinement, culminating in an $R^2$ score improvement to 0.992. While no substantial variations in the trends of feature importance distributions are observed between time steps 50 and 1000, minor adjustments in specific values are exhibited, indicating that the OMT structures are being updated in response to these slight changes. Therefore, it can be inferred that the most significant alterations of the OMT structures during the ENA process occur in the initial time steps, with subsequent iterations leading to a gradual convergence towards optimized solutions. The detailed analysis of the dynamics of feature matrices extracted from the OMT-FNN is presented in Section 7 of the Supplement.

### 4.2.5 Flexibility demonstration of the ENA method

To demonstrate the flexibility of the ENA method, we constrain the material and thickness of the first layer of OMTs to $SiO_2$ and 100 nm. The target structure listed in Table S7 of the Supplement is used. In this experiment, we set $w_3$ as 1.

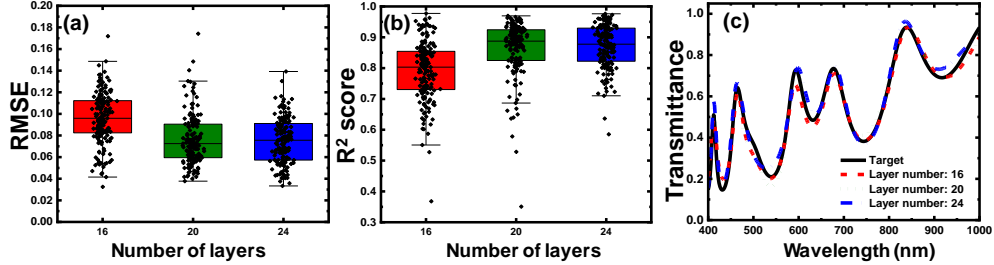

Fig. 10. Box overlap plot for (a) RMSE and (b) $R^2$ score of the constrained OMT solutions with layer numbers of 16, 20, and 24 obtained using the ENA method, and (c) Transmittance spectra for the target structure and those obtained using the ENA method.

Table 5. The constrained OMT structures with layer lengths of 16, 20, and 24, designed using the ENA method. Thickness is reported to one decimal place.

| Layer number | Inverse-designed material configurations | Inverse-designed thicknesses (nm) |
|---|---|---|
| 16 | $SiO_2$/AlN/$Al_2O_3$/MgO/ZnSe/ZnO/$HfO_2$/$SiO_2$/ $Al_2O_3$/$TiO_2$/$Al_2O_3$/ZnSe/$MgF_2$/ZnS/$Al_2O_3$/$TiO_2$ | 99.7/21.5/22.5/71.0/87.3/40.1/97.0/96.3/ 75.6/77.7/41.2/95.7/26.7/76.6/60.4/36.7 |
| 20 | $SiO_2$/$TiO_2$/ZnSe/$HfO_2$/ZnS/MgO/AlN/MgO/ $HfO_2$/ZnSe/MgO/$Si_3N_4$/$TiO_2$/ZnSe/$MgF_2$/$Al_2O_3$/ $HfO_2$/$SiO_2$/ZnS/$MgF_2$ | 99.9/20.4/23.5/97.4/27.7/63.8/44.2/82.1/ 29.0/27.3/67.9/75.1/42.4/64.3/81.4/40.6/ 49.7/56.4/94.8/21.2 |
| 24 | $SiO_2$/$Al_2O_3$/$TiO_2$/ZnSe/$HfO_2$/$Al_2O_3$/AlN/$SiO_2$/ MgO/ZnO/$Si_3N_4$/ZnS/$Si_3N_4$/$SiO_2$/ZnSe/$Al_2O_3$/ $SiO_2$/ZnO/$Al_2O_3$/$MgF_2$/ZnSe/$Al_2O_3$/ZnS/$Al_2O_3$ | 100.0/99.0/65.9/42.5/97.2/33.9/73.2/23.5/ 24.3/97.9/46.6/47.2/21.7/27.9/69.8/40.0/ 71.7/47.1/33.8/75.2/42.6/47.3/21.2/20.7 |

Table 6. The performance metrics of the constrained OMT structures with layer lengths of 16, 20, and 24, designed using the ENA method.

| Layer number | RMSE | $R^2$ score | $|C|$ | $\sigma_t$ | $\sigma_m$ |
|---|---|---|---|---|---|
| 16 | 0.033 | 0.977 | 23 | 24.102 | 3.081 |
| 20 | 0.039 | 0.967 | 61 | 24.775 | 3.110 |
| 24 | 0.033 | 0.976 | 67 | 25.039 | 3.065 |

Figure 10(a) and (b) present box overlap plots illustrating RMSE and $R^2$ score for the constrained OMT solutions with layer numbers of 16, 20, and 24 obtained from the ENA method. Table 5 presents the OMT structures exhibiting the flexibility achieved through the ENA method. The transmittance spectra and performance metrics for constrained OMT structures listed in Table 5 are shown in Fig. 10(c) and Table 6.

Table 5 indicates that the first layers of multilayers with the layer numbers 16, 20, and 24 consist of $SiO_2$ with a thickness of approximately 100 nm. Furthermore, the initial layers of 127 among a total of 151 OMT solutions also conform to the $SiO_2$. Additionally, the overall OMT solutions exhibit a thickness of approximately 100 nm. These results demonstrate the flexibility of the proposed ENA method. The best accuracy reported in Table 6 is inferior to that of the ENA method listed in Table 2, which can be attributed to artificial constraints. Nevertheless, it is important to note that the summation of $|C|$ in Table 6, amounting to 151, exceeds that of the ENA without material loss and the ENA without material redundancy regularization loss presented in Table 2. This means that the diversity of the ENA method, despite applying to artificial constraints, is superior to the ENA without material loss and the ENA without material redundancy loss. This finding indicates that exploring the material configurations is essential to enhance the efficacy of the inverse design method associated with the OMT.

*5. Conclusion*

We proposed the ENA framework, which fulfills six crucial criteria outlined in the AIAID: accuracy, efficiency, diversity, scalability, flexibility, and interpretability. To enhance the scalability of the previous NA method, we presented the OMT-FNN architecture, which facilitates precise predictions of optical properties across various material configurations, layer numbers, and thicknesses. The trained OMT-FNN achieved an RMSE of 0.010 and an $R^2$ score of 0.999. To support the exploration of material configurations within the ENA framework, we introduced a material loss, which consists of boundary loss for materials and material redundancy regularization loss, into the existing NA loss function. Furthermore, to improve the interpretability, we presented a comprehensive formulation of the F-RAM, which is a feature visualization technique.

To validate the effect of the material loss introduced in the proposed ENA method for exploring diverse material configurations, we conducted a comparative analysis involving the ENA without material loss, the ENA without material redundancy regularization loss, and the ENA method. We designated a specific test dataset as a target. The ENA method not only yielded optimal OMT solutions with the highest accuracy, but also provided 82 additional solutions compared to the ENA without material loss. When the material redundancy regularization loss was excluded from the ENA, there was a noted decline in performance, with an increase of 0.021 in RMSE, a decrease of 0.029 in $R^2$ score, and a reduction of 84 in solution counts compared to the ENA. Therefore, the optimization of the material loss within the proposed ENA method facilitates exploring material configurations, leading to a substantial enhancement in both accuracy and diversity.

To demonstrate the performance of the proposed ENA method, we undertook a comparative study involving the Res-GLOnet. We targeted a transmittance spectrum that completely transmits light within the wavelength range of 600 nm to 700 nm while reflecting all other wavelengths. As a result, the ENA method yielded the more favorable OMT solution, achieving an RMSE of 0.111 and an $R^2$ score of 0.912. In comparison, the Res-GLOnet produced the OMT solution with an RMSE of 0.114 and an $R^2$ score of 0.907. Notably, the Res-GLOnet exhibited a greater number of selected OMT solutions, generating 18 additional OMT solutions relative to the ENA method. However, the average standard deviations for thickness and material index of the ENA method were 7.4 times and 50.2 times greater than those of the Res-GLOnet, respectively. This analysis suggested that the actual diversity of the ENA method surpasses that of the Res-GLOnet.

We improved the interpretability of the OMT-FNN via F-RAM. F-RAM was employed for the target structure as well as for the five most accurate OMT structures with a layer number of 20, obtained by the ENA method. Additionally, F-RAM was utilized for the OMT structures with layer numbers of 16 and 24, which exhibited the highest accuracy regarding their respective layer number. Consequently, we demonstrated that the trained OMT-FNN assigns analogous feature importance to OMT structures exhibiting similar optical properties, despite variations in material configurations, layer numbers, and thicknesses. Furthermore, to enhance the interpretability of the ENA method, we implemented F-RAM on the updated OMT structures during the ENA method. The analysis conducted using F-RAM indicated that significant changes in OMT structures predominantly occur during the initial phase, whereas subsequent adjustments to the OMT structures are primarily fine-tuning. This suggests that the F-RAM serves as an effective tool for enhancing the interpretability of the OMT-FNN and the ENA method.

Furthermore, we demonstrated the flexibility of the ENA method by restricting the first layer of OMTs to SiO$_2$ and 100 nm. Although the accuracy of the OMT solutions in the ENA was degraded due to the artificial constraints, the diversity was still superior to the ENA without material loss and the ENA without material redundancy regularization loss. In conclusion, our results demonstrated that the ENA method with F-RAM achieves a reasonable accuracy, efficiency, diversity, scalability, flexibility, and interpretability.

**CRediT authorship contribution statement**

**Sungjun Kim:** Conceptualization, Methodology, Software, Visualization, Writing – original draft, Writing – review & editing. **Jungho Kim:** Conceptualization, Methodology, Visualization, Supervision, Writing – original draft, Writing – review & editing.

**Declaration of competing interest**

The authors declare that they have no known competing financial interests or personal relationships that could have appeared to influence the work reported in this paper.

**Data availability**

Data underlying the results presented in this paper are not publicly available at this time but may be obtained from the authors upon reasonable request.

**Supplemental document**

See Supplement 1 for supporting content.

**Acknowledgments**

This research was in part supported by the Basic Science Research Program (NRF-2021R1F1A1062591) through the National Research Foundation of Korea.

# Interpretable inverse design of optical multilayer thin films based on extended neural adjoint and regression activation mapping: supplement

Sungjun Kim and Jungho Kim[*]

*Department of Information Display, Kyung Hee University, 26 Kyungheedae-ro, Dongdaemun-gu, Seoul 02447, Republic of Korea*

\* Corresponding author.
*E-mail addresses*: sungjunkim@khu.ac.kr (S. Kim), junghokim@khu.ac.kr (J. Kim).

## 1. Hyperparameters used in experiments

**Table S1. Hyperparameter settings.**

| Hyperparameter variables | Options |
|---|---|
| Input dimension ($d_i$) | 24 |
| Output dimension ($d_o$) | 301 |
| Embedding dimension ($d_e$) | 400 |
| The number of layers | 8 |
| Kernel size in CNN | 3 |
| The number of heads in T-E | 8 |
| Optimizer | Adam |
| Learning rate in training forward neural networks | $10^{-4}$ |
| Learning rate in the ENA | $10^{-2}$ |
| Epochs in training forward neural networks | 500 |
| Epochs in ENA | 1001 |
| Training batch size | 500 |
| Validation batch size | 125 |
| Test batch size | 125 |
| Epochs in transfer learning | 100 |
| Training batch size in transfer learning | 100 |
| Validation batch size in transfer learning | 25 |
| Test batch size in transfer learning | 25 |

**Table S2. Materials used for calculation according to the indices.**

| Material index | -1 | 0 | 1 | 2 | 3 | 4 |
|---|---|---|---|---|---|---|
| Material name | None | $MgF_2$ | $SiO_2$ | $MgO$ | $Al_2O_3$ | $HfO_2$ |
| Material index | 5 | 6 | 7 | 8 | 9 | 10 |
| Material name | ZnO | $Si_3N_4$ | AlN | ZnS | $TiO_2$ | ZnSe |

## 2. Performance demonstration of trained forward neural networks

In order to investigate effective architectures that exhibit satisfactory performance, we modified the optical multilayer thin film feature extractor (OMT-FE) layer with the multi-layer perceptron (MLP), convolutional neural network (CNN), and transformer encoder (T-E) layers within the forward neural network for optical multilayer thin film (OMT-FNN) framework, as illustrated in Fig. S1. The masking layer is applied to the CNN and T-E architectures to prevent the contributions of padding rows. Although its application may result in a minor reduction in accuracy, it leads to the enhancement of the interpretability of the model. The OMT-FNN with the MLP layer is designated as the baseline for comparison.

The OMT-FNN architectures with the MLP, CNN, T-E, and OMT-FE layers are trained within a supervised learning framework. We measure and compare performance metrics such as training time, RMSE, and $R^2$ score. In order to enhance the scalability of the OMT-FNN with the OMT-FE layers, applying transfer learning techniques for combinations of polarizations and incidence angles is presented.

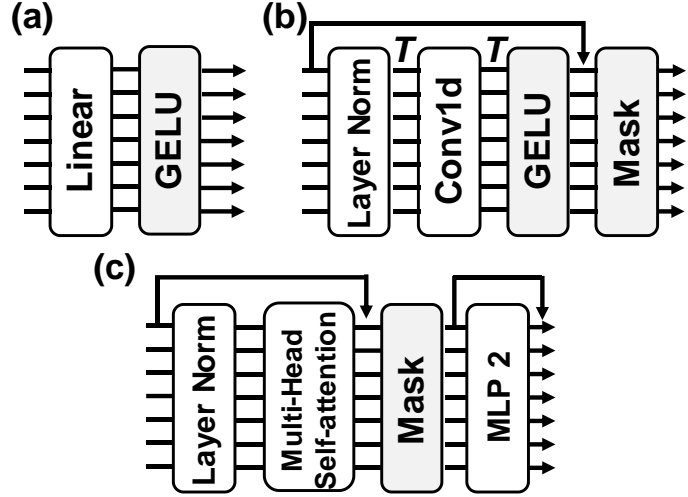

Fig. S1. Deep neural network architectures for comparison: (a) MLP, (b) CNN, (c) T-E.

**Table S3. The performance metrics of trained OMT-FNNs with the MLP, CNN, T-E, and OMT-FE as feature extractors. The values in parentheses next to the training time, RMSE, and $R^2$ score indicate the ratio compared to the baseline, percentage reduction, and percentage improvement, respectively.**

| Feature extractors | Parameters (Million) | Training time (hours) | RMSE | $R^2$ score |
|---|---|---|---|---|
| MLP (Baseline) | 1.24 | 14.5 (1.0) | 0.022 (0.0 %) | 0.993 (0.0 %) |
| CNN | 3.97 | 24.9 (1.7) | 0.017 (22.7 %) | 0.996 (0.3 %) |
| T-E | 10.41 | 60.1 (4.1) | 0.010 (54.5 %) | 0.999 (0.6 %) |
| OMT-FE | 5.13 | 33.2 (2.3) | 0.010 (54.5 %) | 0.999 (0.6 %) |

Table S3 presents the assessed performance metrics of trained OMT-FNNs with the MLP, CNN, T-E, and OMT-FE as feature extractors. Although the OMT-FNN utilizing the MLP layer shows the shortest training time, it exhibits the lowest accuracy compared to other architectures. Nevertheless, the baseline accuracy exceeds 0.99 in $R^2$ score. The MLP layers sequentially operate on linear combinations and activations solely on the embeddings, which effectively facilitates the generation of a robust feature matrix. Conversely, the OMT-FNN incorporating CNN layers experiences a 1.7-fold decrease in efficiency, although it achieves improvements of 22.7 % in RMSE and 0.3 % in $R^2$ score. The CNN layer is responsible for extracting a feature matrix through convolution operations applied exclusively to the OMT layers, thereby enhancing representation quality. The self-attention module within the T-E layer calculates the attention scores across the OMT layers, while the mixing of the embeddings is executed by the MLP 2 layer. This bidirectional mixing leads to a significant enhancement in accuracy, with RMSE and $R^2$ scores improving by 54.5 % and 0.6 %, respectively. However, this improvement comes at the cost of a 4.1-fold reduction in efficiency, a decline that is likely to exacerbate with larger dataset sizes. The OMT-FE layers facilitate the mixing of OMT layers through a straightforward MLP 1 layer, while the embedding mixing via the MLP 2 layer is identical to that of the T-E. By simplifying the self-attention module to an MLP 1 layer, the efficiency of the OMT-FNN with the OMT-FE layers is enhanced compared to the T-E. Notably, this model exhibits comparable accuracy improvement to that of T-E. Thus, the OMT-FNN with the OMT-FE layer acquires a reasonable accuracy and efficiency.

Table S4. The transfer learning results of the OMT-FNN with OMT-FE layers.

|  | s-pol 20° | s-pol 40° | s-pol 60° | p-pol 20° | p-pol 40° | p-pol 60° |
| --- | --- | --- | --- | --- | --- | --- |
| Training time (hours) | 1.3 | 1.3 | 1.3 | 1.3 | 1.3 | 1.3 |
| RMSE | 0.009 | 0.010 | 0.016 | 0.007 | 0.006 | 0.005 |
| $R^2$ score | 0.999 | 0.999 | 0.997 | 0.999 | 0.999 | 0.999 |

To incorporate the prediction of the optical properties according to the incidence angles and polarization, we employ transfer learning, noting that the dataset employed for this purpose comprises only one-tenth of the original optical multilayer thin film (OMT) dataset size. The results of the transfer learning application on the OMT-FNN with the feature extractor of the OMT-FE layer are presented in Table S4. The overall performance of the trained OMT-FNN indicates a commendable level of accuracy and efficiency.

## 3. Comparison of OMT-FNN to OL-Transformer

Table S5. The performance metrics of the trained OL-Transformer. The values in parentheses next to the training time, RMSE, and $R^2$ score indicate the ratio compared to the baseline, percentage reduction, and percentage improvement, respectively.

| Comparative model | Parameters (Million) | Training time (hours) | RMSE | $R^2$ score |
| --- | --- | --- | --- | --- |
| OL-Transformer | 10.91 | 58.2 (4.0) | 0.004 (81.8 %) | 1.000 (0.7 %) |

The performance of the trained OMT-FNN is compared to that of the OL-Transformer (Ma et al., 2023). As shown in Table S5, the OL-Transformer achieves the highest accuracy among the models evaluated, whereas it exhibits similar efficiency to the OMT-FNN with T-E layers. Even though the OL-Transformer possesses a greater number of parameters compared to the OMT-FNN with T-E layers, the marginally increased training time of the OMT-FNN with T-E layers can be attributed to the application of the masking layer. The trained OL-Transformer processes material and thickness as a word, which tackles the thickness to be discrete, set at 1 nm in this study. In contrast, the proposed OMT-FNN model covers the material and thickness as continuous real values, thereby enabling its application as an artificial intelligence-assisted inverse design (AIAID) method utilizing backpropagation, such as the extended neural adjoint (ENA) method. Additionally, our analysis indicates that OMT-FNN with the OMT-FE achieves satisfactory accuracy and is superior to the OL-Transformer in terms of efficiency.

## 4. Inverse-designed OMT structures for Section 4.2.1

**Table S6. Two OMT structures obtained using Res-GLOnet and ENA.**
**Thickness is reported to one decimal place.**

| Model | Inverse-designed material configurations | Inverse-designed thicknesses (nm) |
|---|---|---|
| Res-GLOnet | ZnSe/Si$_3$N$_4$/MgF$_2$/ZnSe/MgF$_2$/ZnS/MgF$_2$/ZnS/ MgO/ZnSe/AlN/SiO$_2$/TiO$_2$/MgF$_2$/SiO$_2$/ZnS/ TiO$_2$/MgF$_2$/MgF$_2$/ZnSe/MgF$_2$/MgF$_2$/ZnSe/TiO$_2$ | 92.2/69.7/96.1/51.0/99.1/33.2/99.6/60.7/ 98.9/49.4/99.2/99.3/90.1/53.8/46.4/89.3/ 28.4/97.4/33.2/92.5/89.3/27.3/24.1/93.2 |
| ENA | ZnO/Si$_3$N$_4$/MgF$_2$/ZnSe/MgF$_2$/ZnSe/AlN/ZnS/ MgF$_2$/ZnSe/Al$_2$O$_3$/MgF$_2$/ZnO/MgO/MgF$_2$/MgO/ MgF$_2$/TiO$_2$/MgF$_2$/TiO$_2$/MgF$_2$/TiO$_2$/ZnSe/AlN | 53.6/98.3/99.8/95.1/99.7/43.6/28.3/54.5/ 99.9/86.0/20.0/99.2/78.4/23.6/68.2/38.1/ 98.4/56.4/77.4/45.3/97.2/31.4/94.5/20.5 |

The OMT structures generated by the Res-GLOnet (Jiang and Fan, 2020) and the ENA method are detailed in Table S6. The transmittance spectra for OMT structures listed in Table S6 are shown in Fig. 5(c).

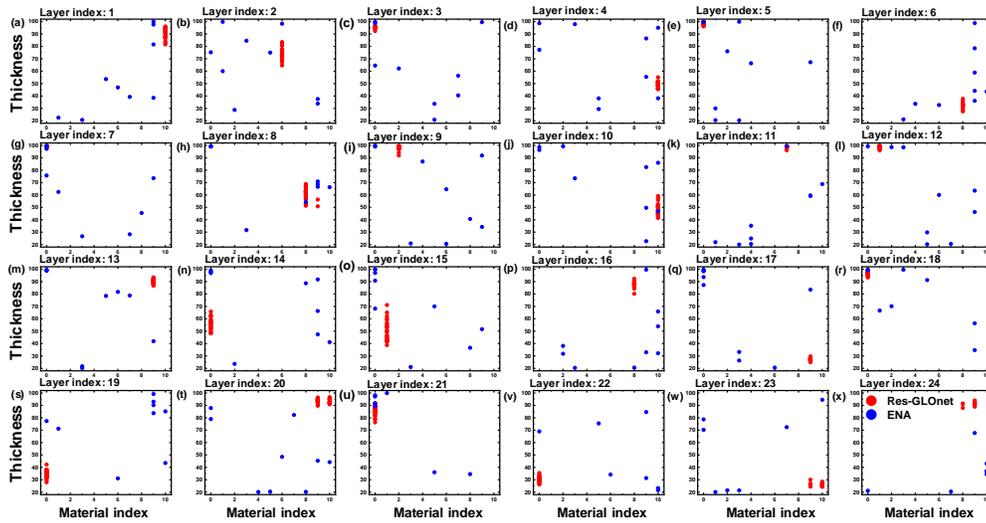

Fig. S2. Scatter plots for material index and thickness of the OMT solutions according to layer index derived from the Res-GLOnet (red dot) and the ENA (blue dot).

Figure S2 shows scatter plots for the material index and thickness of the OMT solutions according to layer index obtained by the Res-GLOnet and the ENA. The solutions generated by the Res-GLOnet exhibit a concentration around a singular material configuration at a specific thickness combination, in contrast to the more dispersed solutions produced by the ENA method.

## 5. Target structure and inverse-designed structures from Section 4.2.2 to Section 4.2.3

**Table S7. Target OMT structure and the three OMT structures obtained using the ENA without material loss, the ENA without material redundancy regularization loss, and the ENA. Thickness is reported to one decimal place.**

|  | Material configurations | Thicknesses (nm) |
|---|---|---|
| Target structure | $Al_2O_3$/$Si_3N_4$/ZnS/MgO/ZnSe/ZnO/ZnS/ZnSe/ZnO/ZnS/$MgF_2$/$SiO_2$/ZnSe/$MgF_2$/$HfO_2$/AlN/$SiO_2$/AlN/$Al_2O_3$/$SiO_2$ | 32/42/61/30/73/23/74/69/93/71/57/25/36/32/54/51/67/26/81/23 |
| Model | Inverse-designed material configurations | Inverse-designed thicknesses (nm) |
| ENA w/o $L_m$ | AlN/ZnSe/$Si_3N_4$/AlN/ZnO/$MgF_2$/$HfO_2$/$TiO_2$/ZnSe/MgO/AlN/$SiO_2$/$MgF_2$/ZnSe/MgO/$Al_2O_3$/AlN/$HfO_2$/ZnO/MgO | 58.2/37.6/60.6/47.1/30.8/74.8/41.6/70.3/86.9/68.5/25.5/23.7/94.1/36.9/28.4/20.8/39.8/81.1/73.4/37.6 |
| ENA w/o $L_m^{(r)}$ | ZnO/$SiO_2$/$TiO_2$/$SiO_2$/ZnSe/$Al_2O_3$/MgO/ZnSe/ZnS/ZnO/$TiO_2$/ZnSe/ZnO/$Al_2O_3$/MgO/ZnO/$HfO_2$/$TiO_2$/ZnO/$SiO_2$ | 49.4/52.7/67.8/57.9/90.4/30.2/20.9/90.9/31.5/77.1/84.6/87.7/32.3/77.0/69.8/91.3/92.4/95.6/88.0/98.2 |
| ENA | MgO/$Al_2O_3$/ZnO/$MgF_2$/$TiO_2$/ZnSe/$MgF_2$/ZnO/$Si_3N_4$/ZnO/ZnSe/ZnO/MgO/$Al_2O_3$/$SiO_2$/$Al_2O_3$/MgO/ZnS/$TiO_2$/$Al_2O_3$ | 30.9/31.0/64.0/91.0/20.2/24.8/75.3/55.0/20.1/27.9/76.4/99.3/35.9/21.1/78.4/63.7/62.9/85.3/55.2/67.6 |

Table S7 delineates the target OMT structure alongside the OMT structures exhibiting the highest accuracy achieved through the ENA without material loss, the ENA without material redundancy regularization loss, and the ENA method.

**Table S8. The five most accurate OMT structures based on accuracy with a layer number of 20, designed using the ENA method. Thickness is reported to one decimal place.**

| Rank | Inverse-designed material configurations | Inverse-designed thicknesses (nm) |
|---|---|---|
| Top 1 | MgO/$Al_2O_3$/ZnO/$MgF_2$/$TiO_2$/ZnSe/$MgF_2$/ZnO/$Si_3N_4$/ZnO/ZnSe/ZnO/MgO/$Al_2O_3$/$SiO_2$/$Al_2O_3$/MgO/ZnS/$TiO_2$/$Al_2O_3$ | 30.9/31.0/64.0/91.0/20.2/24.8/75.3/55.0/20.1/27.9/76.4/99.3/35.9/21.1/78.4/63.7/62.9/85.3/55.2/67.6 |
| Top 2 | $SiO_2$/ZnS/ZnSe/ZnS/MgO/$TiO_2$/$SiO_2$/ZnSe/ZnO/$Si_3N_4$/AlN/ZnO/$Si_3N_4$/$HfO_2$/$TiO_2$/ZnO/$Al_2O_3$/ZnS/$SiO_2$/$HfO_2$ | 32.6/50.3/44.3/97.9/49.1/46.1/78.1/59.6/75.7/22.2/24.6/25.3/73.2/73.0/68.9/40.6/32.8/42.5/48.9/27.8 |
| Top 3 | MgO/ZnSe/$TiO_2$/MgO/$MgF_2$/ZnS/$Al_2O_3$/$HfO_2$/$TiO_2$/$Al_2O_3$/AlN/$Al_2O_3$/ZnSe/MgO/AlN/$MgF_2$/$TiO_2$/$Si_3N_4$/$Al_2O_3$/$SiO_2$ | 69.3/64.7/84.9/21.1/20.3/31.6/61.8/25.0/99.3/34.0/59.2/63.8/34.9/68.8/79.2/95.5/39.4/22.5/27.4/44.3 |
| Top 4 | MgO/$MgF_2$/ZnO/ZnS/ZnO/ZnSe/ZnO/$MgF_2$/MgO/$Al_2O_3$/$HfO_2$/$TiO_2$/$Al_2O_3$/$Si_3N_4$/ZnSe/$MgF_2$/$Al_2O_3$/$TiO_2$/$Si_3N_4$/ZnS | 43.4/22.3/36.8/29.4/65.7/99.0/68.3/75.6/60.4/88.6/28.3/46.5/23.1/22.1/38.2/48.1/24.0/58.7/53.3/93.8 |
| Top 5 | $MgF_2$/ZnSe/$TiO_2$/ZnSe/$Al_2O_3$/MgO/ZnSe/$Al_2O_3$/$MgF_2$/$TiO_2$/ZnO/$Si_3N_4$/$HfO_2$/$Al_2O_3$/$Si_3N_4$/MgO/$Al_2O_3$/$TiO_2$/$MgF_2$/$Al_2O_3$ | 70.8/46.2/26.5/23.4/98.1/99.4/20.8/28.3/21.3/96.9/22.8/20.8/98.8/33.4/61.1/23.2/55.8/48.9/93.9/79.2 |

Table S8 shows the five most accurate OMT structures with a layer number of 20, obtained by the ENA method.

**Table S9. The OMT structures with layer numbers of 16, 20, and 24, designed using the ENA method. Thickness is reported to one decimal place.**

| Layer numbers | Inverse-designed material configurations | Inverse-designed thicknesses (nm) |
|---|---|---|
| 16 | $Si_3N_4$/$TiO_2$/ZnS/ZnO/$MgF_2$/AlN/$TiO_2$/ZnSe/MgO/ZnO/$MgF_2$/ZnSe/AlN/$Al_2O_3$/ZnSe/$HfO_2$ | 64.6/42.4/99.2/99.1/93.3/72.8/71.9/35.3/29.4/76.5/99.7/57.0/21.0/34.1/47.6/70.6 |
| 20 | MgO/$Al_2O_3$/ZnO/$MgF_2$/$TiO_2$/ZnSe/$MgF_2$/ZnO/$Si_3N_4$/ZnO/ZnSe/ZnO/MgO/$Al_2O_3$/$SiO_2$/$Al_2O_3$/MgO/ZnS/$TiO_2$/$Al_2O_3$ | 30.9/31.0/64.0/91.0/20.2/24.8/75.3/55.0/20.1/27.9/76.4/99.3/35.9/21.1/78.4/63.7/62.9/85.3/55.2/67.6 |
| 24 | $MgF_2$/AlN/$Si_3N_4$/$TiO_2$/$SiO_2$/ZnSe/$MgF_2$/ZnO/$MgF_2$/$HfO_2$/$SiO_2$/$MgF_2$/$TiO_2$/$HfO_2$/MgO/ZnSe/$TiO_2$/$Si_3N_4$/$HfO_2$/$TiO_2$/AlN/MgO/$MgF_2$/$Al_2O_3$ | 27.6/58.3/69.8/48.5/37.6/34.5/91.1/26.3/20.3/27.2/77.8/79.0/26.2/84.3/20.9/80.7/44.1/55.3/65.5/58.1/27.1/95.0/70.4/20.4 |

Table S9 shows the OMT structures with layer numbers of 16, 20, and 24, obtained by the ENA method.

## 6. Regression activation mapping results of forward neural networks for optical multilayer thin films

We apply regression activation mapping of OMT-FNN (F-RAM) to 100 test data with layer numbers of 4, 8, 12, 16, 20, and 24. These results are presented in Figures S3, S4, S5, S6, S7, and S8.

Our analysis is focused on the F-RAMs for OMTs of a layer number of 20, presented in Figure S7. We can discern the importance of feature vectors through the F-RAM in predicting the optical properties. Notably, F-RAMs calculated by OMT-FNNs do not typically provide the importance of feature vectors, with their indices ranging from 21 to 24, where they correspond to padding layers. This suggests that the proposed OMT-FNN treats features in accordance with the layer number of the OMTs. Furthermore, the patterns observed in F-RAMs vary, reflecting the distinct feature matrices extracted in accordance with the different models employed. Based on established domain knowledge regarding OMTs, it is evident that each OMT layer plays a crucial role in determining optical properties. However, the OMT-FNN with the MLP layers emphasizes the significance of individual layers. Conversely, the OMT-FNN with the CNN, T-E, and OMT-FE layers distributes importance across all layers. Specifically, for the CNN and T-E layers, the importance is uniformly distributed, while the OMT-FE layer exhibits a slight concentration of importance in a particular layer, albeit not to a significant extent, maintaining an overall distribution of relevance.

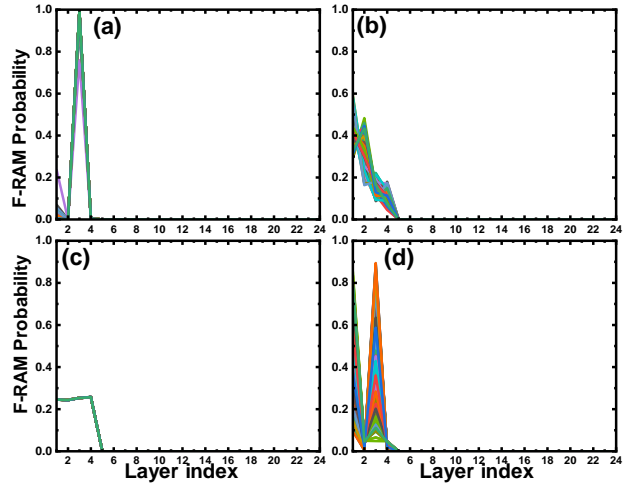

Fig. S3. The F-RAMs for 100 test data with a layer number of 4 of the OMT-FNN with (a) the MLP, (b) CNN, (c) T-E, and (d) OMT-FE layers

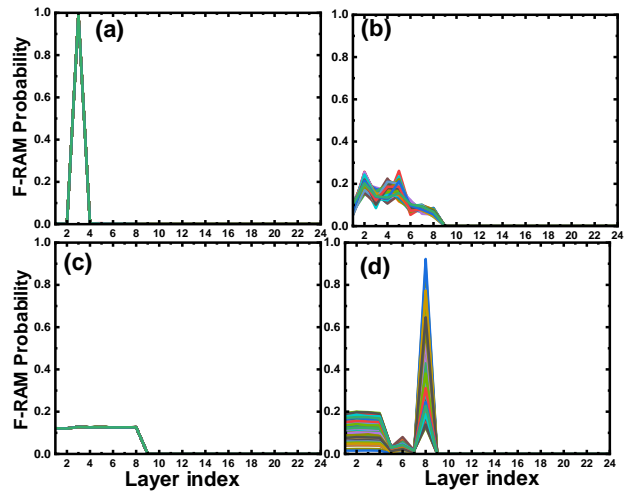

Fig. S4. The F-RAMs for 100 test data with a layer number of 8 of the OMT-FNN with (a) the MLP, (b) CNN, (c) T-E, and (d) OMT-FE layers

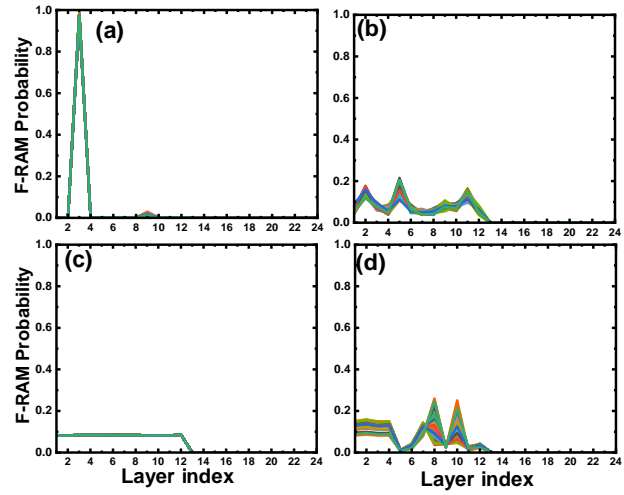

Fig. S5. The F-RAMs for 100 test data with a layer number of 12 of the OMT-FNN with (a) the MLP, (b) CNN, (c) T-E, and (d) OMT-FE layers

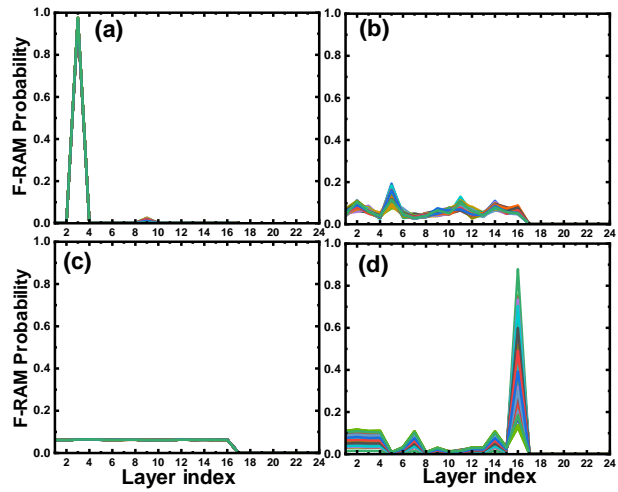

Fig. S6. The F-RAMs for 100 test data with a layer number of 16 of the OMT-FNN with (a) the MLP, (b) CNN, (c) T-E, and (d) OMT-FE layers

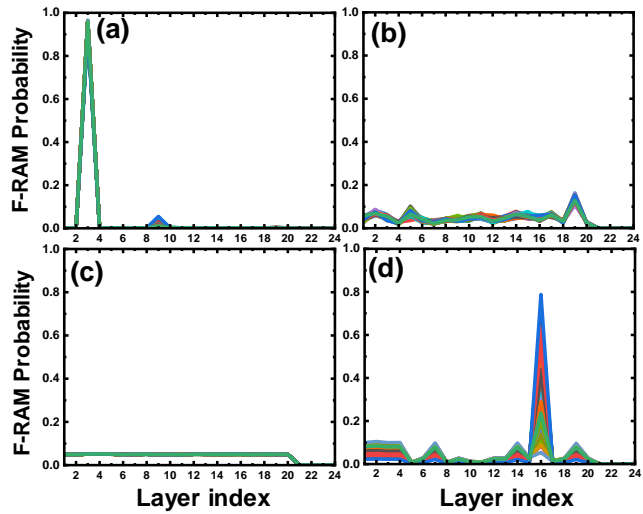

Fig. S7. The F-RAMs for 100 test data with a layer number of 20 of the OMT-FNN with (a) the MLP, (b) CNN, (c) T-E, and (d) OMT-FE layers

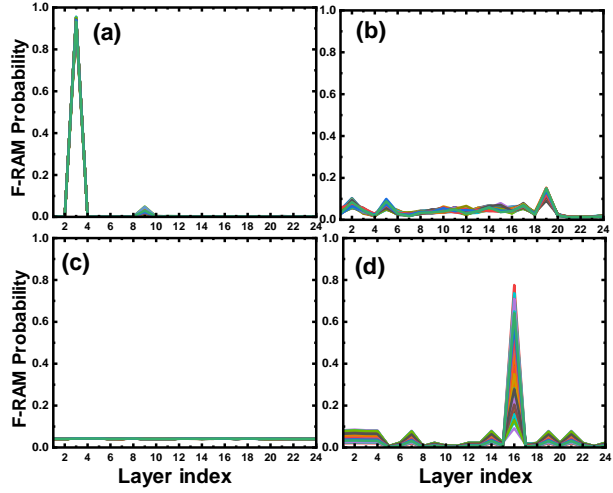

Fig. S8. The F-RAMs for 100 test data with a layer number of 24 of the OMT-FNN with (a) the MLP, (b) CNN, (c) T-E, and (d) OMT-FE layers

## 7. Explanation of the dynamics of feature matrices through F-RAM

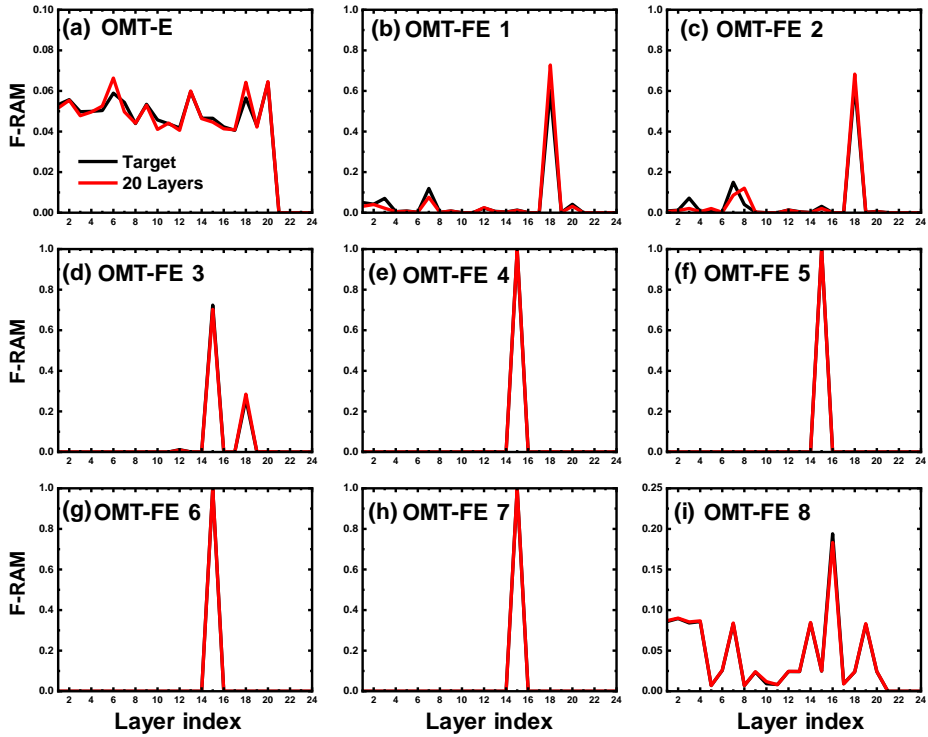

Fig. S9. F-RAMs according to the layer depth of the OMT-FNN for the target structure and the OMT structure with a layer length of 20, listed in Table S9.

As shown in Fig. S9(i), the F-RAMs corresponding to the target structure and the OMT structure of a layer number of 20 listed in Table S9 exhibit a high degree of similarity. To further elucidate the evolution of F-RAM in relation to layer depth, Figure S9 is presented. Figure S9(a) illustrates the F-RAMs for the embedding matrices obtained by the OMT-E layer of the two OMT structures. Despite variation in material and thickness across the layers, the resulting F-RAMs remain comparable. The initial OMT-FE layer induces a significant transformation in the embedding matrices, thereby altering the trend of the F-RAMs. From the second to the seventh OMT-FE layers, the extracted feature matrices undergo fine-tuning, resulting in only minor modifications to the F-RAMs. Notably, from the third OMT-FE layer onward, the F-RAM converge to identical values. The final OMT-FE layer redistributes the elements of the feature matrix concentrated within 15 rows, while the F-RAMs of the two OMT structures continue to align. The first and final OMT-FE layers effectuate substantial transformations in the feature matrices, whereas the intermediate OMT-FE layers induce only slight alterations. Consequently, the dynamics of the feature matrices and the role of components of the OMT-FNN can be effectively characterized through the F-RAM.